\documentclass[letterpaper,10pt,conference]{IEEEtran}

\usepackage{amsmath,amssymb,mathrsfs,bm}
\usepackage{caption,subfigure}
\usepackage{graphicx,color,xspace}
\usepackage{algorithm,url}
\usepackage{mathtools} 

\usepackage[noend]{algpseudocode}
\usepackage{float}
\makeatletter
\def\BState{\State\hskip-\ALG@thistlm}
\makeatother

\usepackage{dsfont}
\usepackage{tikz}

\usepackage{stfloats} 

\usepackage{accents}

\newcommand{\oprocendsymbol}{\hbox{$\bullet$}}
\newcommand{\oprocend}{\relax\ifmmode\else\unskip\hfill\fi\oprocendsymbol}

\newcommand{\longthmtitle}[1]{\mbox{}\textup{(#1):}}

\newtheorem{theorem}{Theorem}[section]

\newtheorem{remark}[theorem]{Remark}

\newcommand{\margind}[1]{\marginpar{\color{blue}\tiny\ttfamily#1}}

\renewcommand{\baselinestretch}{.99}

\newcommand{\bmse}{{\rm{BMSE}}}

\newcommand{\sigmasignal}{\sigma_s}

\newcommand{\nrx}{{n}}
\newcommand{\mtx}{{m}}

\newcommand{\idtx}{d^\text{\scalebox{0.9}{Tx}}}
\newcommand{\idrx}{d^\text{\scalebox{0.9}{Rx}}}

\newcommand{\dtx}{d^\text{\scalebox{0.9}{Tx}}}
\newcommand{\drx}{d^\text{\scalebox{0.9}{Rx}}}
\newcommand{\dvirt}{d^\text{\scalebox{0.9}{Virt}}}

\newcommand{\dtxm}[1]{\idtx_{#1}}
\newcommand{\drxn}[1]{\idrx_{#1}}

\newcommand{\noise}{w}
\newcommand{\sigmanoise}{\sigma}

\newcommand{\mindistTx}{s^\text{\scalebox{0.9}{Tx}}}
\newcommand{\mindistRx}{s^\text{\scalebox{0.9}{Rx}}}
\newcommand{\rightextremeTx}{U^\text{\scalebox{0.9}{Tx}}}
\newcommand{\leftextremeTx}{L^\text{\scalebox{0.9}{Tx}}}
\newcommand{\rightextremeRx}{U^\text{\scalebox{0.9}{Rx}}}
\newcommand{\leftextremeRx}{L^\text{\scalebox{0.9}{Rx}}}

\newcommand{\costdeltau}{f_{\Delta u}}
\newcommand{\setdeltau}{\mathcal{S}_{\Delta u}}

\newcommand{\deltau}{\Delta u}
\newcommand{\deltauone}{\Delta {u_1}}
\newcommand{\deltaul}{\Delta {u_l}}
\newcommand{\deltaudesired}{\Delta {u_\text{Desired}}}

\newcommand{\wwb}{{\rm{WWB}}}

\newcommand{\tp}{^{\text{\scalebox{0.8}{$\top$}}}}

\newcommand{\defin}{:=}

\newcommand{\until}[1]{\{1,\dots,#1\}}

\newcommand{\map}[3]{#1:#2 \rightarrow #3}

\newcommand{\realnonnegative}{{\mathbb{R}}_{\ge 0}}
\newcommand{\real}{{\mathbb{R}}}

\newcommand{\complex}{{\mathbb{C}}}

\newcommand{\norm}[1]{\|#1\|_{\text{\scalebox{1}{$2$}}}}

\newcommand{\absolute}[1]{|#1|}

\begin{document}





\title{Constrained sparse array optimization and performance evaluation via Compressive Sensing}


\title{Sparse Array design via constrained optimization using the Weiss-Weinstein Bound}

\title{Constrained optimal design of automotive radar arrays using the Weiss-Weinstein Bound}
\title{Constrained optimal arrays using the Weiss-Weinstein Bound}
\title{Constrained optimal arrays and performance evaluation with Compressive Sensing}

\title{Constrained optimal design and performance evaluation of automotive radar arrays}
\title{Constrained optimal design and DoA estimation performance of automotive radar arrays}
\title{Constrained optimal design of automotive radar
	arrays using the Weiss-Weinstein Bound}


\author{\IEEEauthorblockN{Mar\'ia A. Gonz\'alez-Huici}
	\and
\IEEEauthorblockN{David Mateos-N\'u\~nez}
	\and
\IEEEauthorblockN{Christian Greiff}
	\and
\IEEEauthorblockN{Renato Simoni}
\thanks{This work has been partially supported by the BMBF project Radar4FAD and motivated by a collaboration with Infineon Technologies AG.}
\thanks{The authors would like to thank Jorge Jacome-M\'u\~noz, Aitor Correas-Serrano, Fabio Giovanneschi, and Stefan Br\"uggenwirth for useful discussions.}
}

\IEEEoverridecommandlockouts

\maketitle

\begin{abstract}
We propose a design strategy for optimizing antenna positions in linear arrays for far-field Direction of Arrival (DoA) estimation of narrow-band sources in collocated MIMO radar. Our methodology allows to consider any spatial constraints and number of antennas, using as optimization function the Weiss-Weinstein bound formulated for an observation model with random target phase and known SNR, over a pre-determined Field-of-View (FoV). Optimized arrays are calculated for the typical case of a 77GHz MIMO radar of 3Tx and 4Rx channels. Simulations demonstrate a performance improvement of the proposed arrays compared to the corresponding uniform and minimum redundancy arrays for a wide regime of SNR values.

\end{abstract}


\section{Introduction}

Array design for DoA estimation in automotive MIMO radar
 demands approaches that constrain the number of antennas and the space available, aimed at improving performance while keeping low complexity and costs.
 Of particular interest for next generation chip design~\cite{RF-CW-SS-SS-HJ-AS:09} 
 are the optimization of antenna positions of non-uniform linear arrays and the associated performance of high-resolution DoA estimation algorithms.




The array factor motivated pioneering work on non-uniform array design and array current synthesis (beamforming) 
through inversion techniques and dynamic programming~\cite{HU:60,MS-GN-JS:64}, mainly using metrics such as Mainlobe Width (MLW) and Sidelobe Levels (SSL). The problem of element spacing synthesis has been acknowledged to be a harder optimization problem than current/weight synthesis~\cite{BPK-GRN:99};
indeed, it requires global optimization, like particle swarm optimization~\cite{MMK-CGC:05,CMS-RF-CW-AS:09,AR:12}. Some recent works have also dealt with the iterative search of both antenna locations and weights~\cite{WR-LX-JL-PS:11}. 
%
%
%
%
%
The main drawback of the array factor as an optimization principle is that the desired trade-off between MLW and SSL is selected \textit{ad hoc}, without regard for the SNR.

%


Some classical array concepts are formulated in terms of diversity of the co-array, like Minimum Redundancy Arrays (MRA)~\cite{DAL-IHS-IGT:93}. 
Finding these arrays requires exhaustive search but some construction methods have been developed for SIMO arrays and extended to the MIMO case, like MRAs based on Cyclic Difference Sets~\cite{JD-QL-WG:09} and nesting procedures~\cite{AK-JG-JD:14}. The main disadvantage is that array size is dictated by the number of antennas, and vice versa. Recent work~\cite{RR-VK:17} reformulates a construction method for a class of SIMO MRAs called Wichmann arrays as a function of desired aperture, but does not allow to constraint simultaneously the number of elements. 

Bayesian approaches have been considered more recently using information-theoretic metrics,
 %
 like the mutual information between measurements and the source angle
in terms of array positions~\cite{GS-SJ-GWW:17}, 
or 
 bounds on the Bayesian Mean-Squared Error (BMSE), which are an indicator of the achievable performance of any estimator and thus  quantify the information that can be extracted from the scene for a candidate non-uniform array. The work~\cite{UO-RLM:05} optimizes the Bayesian and Expected Cram\'er-Rao bounds (BCRB, ECRB)~\cite{HLVT-KLB:13}, and the Expected Fisher Information Matrix, which are related to the MLW~\cite{HM:92} and are better suited for design at high SNR where large MSE errors due to sidelobe ambiguity are not predominant.  Other bounds, like the Weiss-Weinstein bound (WWB)~\cite{AW-EW:85,AR-PF-PL-CDR-AN:08} and the Ziv-Zakai bound (ZZB)~\cite{DK-KLB:10} take into account large estimation errors which occur frequently below a certain SNR due to array sidelobes (``threshold effect")~\cite{NDT-AR-RB-SM-PL:12}. These large errors are underestimated by the BCRB.
On the other hand, a widely used modeling assumption for the construction of the WWB and related bounds (called unconditional model~\cite{DTV-AR-RB-SM:14})
considers target signals distributed as complex Gaussians with zero mean~\cite{FA:01,NDT-AR-RB-SM-PL:12}, which implies that the SNR follows a Rayleigh distribution and thus an excessive emphasis on low SNR values is made, limiting the design choices regarding the SNR of interest. 
The work~\cite{DTV-AR-RB-SM:14} includes several explicit WWB derivations for DoA estimation, particularly under the aforementioned unconditional model and also for the case where both the SNR and the phase are assumed known.
%

%
In this work, we consider an alternative model that 
selects a specific SNR value of interest, and still takes into account the random nature of the target signal phase.
%
%
%
Then, we provide an analytical  WWB for this model as a function of the so-called test points for the target phase and the DoA for a given FoV.”
%
In addition, we present an optimization strategy that permits to include constraints on available space and antenna separation, and use this method to design arrays relevant for automotive radar chips operating at $77$GHz. We then validate and compare the performance of these arrays using a standard open-source sparse reconstruction algorithm for angular estimation.

\section{Sparse array design}\label{sec:sparse-array-design}
This section introduces the observation model for DoA estimation, followed by the optimization metric for array design and the constraints, and finally describes the optimization strategy and presents some examples of optimized 3x4 arrays.
%
\subsection{Signal model for DoA estimation}
 
\begin{figure}[H]
	\hspace*{0.90cm}
	\includegraphics[width=0.77\linewidth]{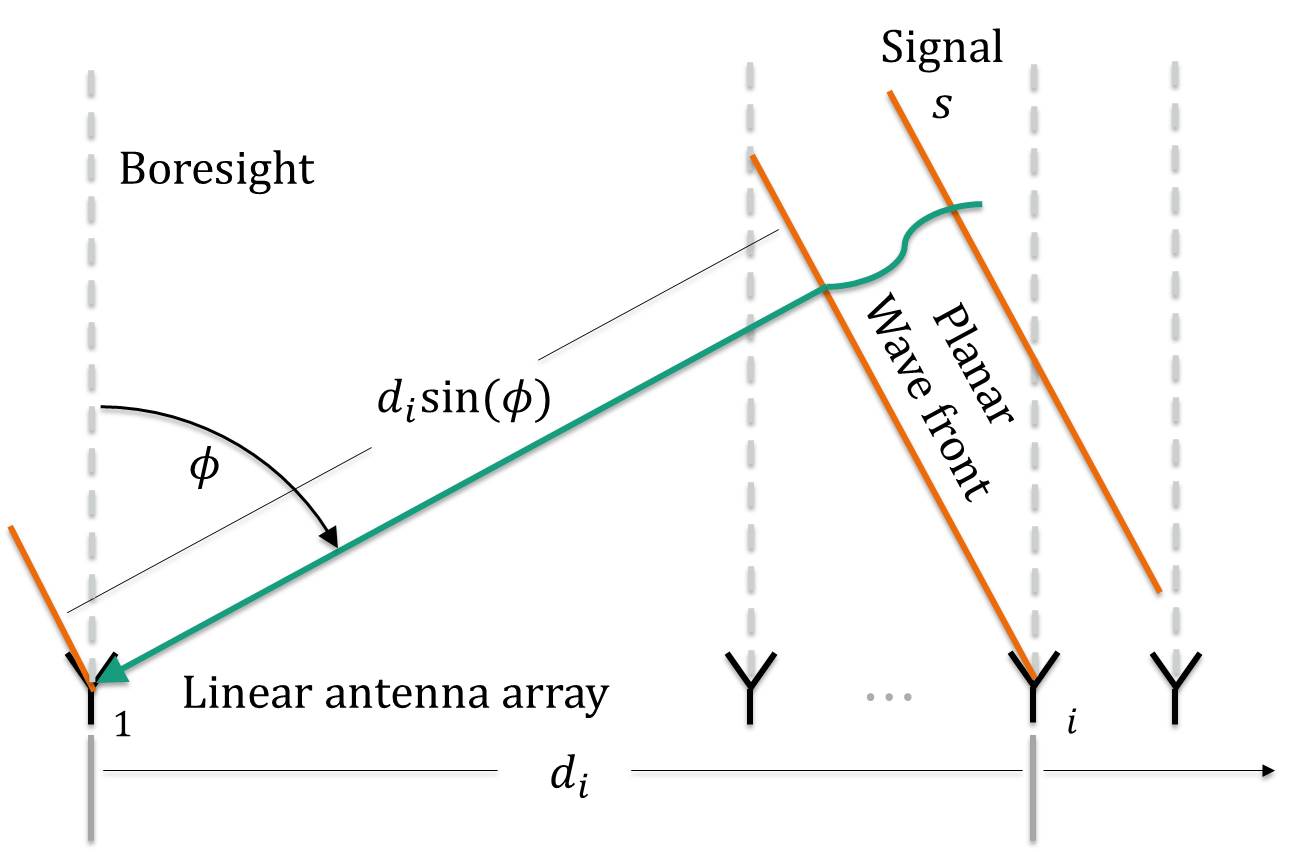}
	\caption{Far-field model for narrow-band sources. Phase shift is proportional to the time it takes for the planar wavefront to reach successive array elements. The antenna positions can correspond to virtual combinations of transmitters and receivers.
		%
	}
	\label{fig:array-diagram}
\end{figure}

Our model for $1$-snapshot DoA estimation of a single far-field source 
 for a collocated MIMO linear array is given by
 \begin{align} \label{eq:measurement_model}
  y = s e^{ik(\dtx\oplus\;\drx) u}  + \noise.
 \end{align}
%
Here, the transmitter and receiver locations are denoted by 
$\dtx\defin [\dtxm{1}, \cdots, \dtxm{\mtx}]\in\real^{\mtx}$ 
and 
$\drx\defin [\drxn{1}, \cdots, \drxn{\nrx}]\in\real^{\nrx}$ (with respect to some reference point), and the positions of the virtual elements $\dvirt\in\real^N$, $N=\mtx\nrx$, are given by
 \begin{align} \label{eq:virtual-positions}
\dvirt=\dtx\oplus\;\drx\defin [\dtxm{1}+\drxn{1}, \dtxm{1}+\drxn{2},\cdots, \dtxm{\mtx}+\drxn{\nrx}] .
\end{align}
We also define $k \defin \frac{2\pi}{\lambda}$ as the wavenumber; $u=\sin(\phi)\in [-1, 1]$ as the parameter of interest, where $\phi$ is the angle (DoA) with respect to boresight, cf. Fig.~\ref{fig:array-diagram}; $s = \absolute{s} e^{i\varphi}\in \complex$ represents the target signal; and $\noise\sim \mathcal{N}_\complex(0, \sigmanoise^2I_N)$ denotes white measurement noise with variance $\sigmanoise^2$. 
%


\subsection{Design cost function}
We wish to design arrays optimized for angle estimation in a FoV $[u_1, u_2]\subset [-1, 1]$ by identifying transmitter and receiver positions $\dtx$ and $\drx$ that minimize the Bayesian Mean-Squared Error (BMSE) of any estimator $\hat{u}\equiv\hat{u}(y)$, 
\begin{align*}
\bmse(\hat{u}) = \frac{1}{\Delta u}\int_{u\in[u_1,u_2]}\int_{y\in\complex^N}(\hat{u}(y)- u)^2 p(y| u) d y d u ,
\end{align*}
where $p(y| u)$ is the \textit{likelihood} of the measurement given $u$, 
%
and we have specified the \textit{prior} belief of $u$ as  uniformly distributed over the FoV length $\Delta u \defin u_2-u_1$.
The BMSE is in general numerically expensive, so we follow
the common practice of replacing it by a tight lower bound whose computation is more convenient. To this end, 
we use the WWB for model~\eqref{eq:measurement_model} where the target phase~$\varphi$ is uniformly  distributed over $[0,  2\pi]$, and $\absolute{s}$ is assumed deterministic and known.\footnote{Our derivation of the WWB uses the MSE also for the phase, although a cyclic cost like the Mean-Cyclic-Error (MCE)~\cite{EN-TR-JT:16} is a good alternative.} This allows to select the regime of SNR values of interest.
The expression of the WWB for this model is derived in the Appendix (see also~\cite{CG-DMN-MGH-SB:18}).
%
 Define
 the SNR as $c\defin\absolute{s}^2/\sigmanoise^2$, and the array factor scaled by the number of antennas as 
 $B(h_u) \defin \frac{1}{N} \sum_{n = 1}^N e^{i k d_nh_u}$, 
 for $d=\dvirt$ in~\eqref{eq:virtual-positions}. Then the family of bounds $\wwb (h_u,h_\varphi)$, parametrized by the so-called test points $(h_u,h_\varphi)\in [0, \Delta u ]\times [-2 \pi,2 \pi]
 $\footnote{For the optimization we selected  $ [10^{-4}, \Delta u]\times [-2 \pi,2 \pi]$.}, are given in~\eqref{eq:randomphaseWWB}. We propose the following cost function for array optimization for a given FoV and SNR,

%
\begin{figure*}[t]
	\small
	\begin{align}\label{eq:randomphaseWWB}
	&\wwb (h_u,h_\varphi)=
	\frac{h_u^2 
		(2\pi - \absolute{h_\varphi})^2
		(\Delta u - \absolute{ h_u})^2 \exp(-cN[1-\Re\{e^{ih_\varphi} B(h_u)\}])}
	{ 2(2\pi\Delta u) \big[(2\pi-\absolute{ h_\varphi})
		(\Delta u -\absolute{ h_u})   -
		\max(0, 2\pi-2\absolute{ h_\varphi})
		\max(0, \Delta u-2\absolute{ h_u})
		\exp(-\frac{cN}{2} [1-\Re \{e^{i2 h_\varphi}B(2h_u)\}]) \big] } .
	\end{align}
	\normalsize
\end{figure*}

\begin{align}\label{eq:opt-objective-design}
\costdeltau(\dtx, \drx)\defin
\sup_{\substack{h_u\in [ 10^{-4}, \Delta u]\\ h_\varphi\in [-2 \pi ,2 \pi]}}
\wwb(h_u,h_\varphi),
\end{align}
which requires,  for the evaluation \textit{of each candidate array}, a global optimization problem of its own to produce the associated tightest bound of the BMSE, $\costdeltau(\dtx, \drx)\le \bmse(\hat{u}).$ With this metric, sidelobes that produce large errors, i.e., far from the mainlobe, are penalized more, while in  practice, estimated targets outside a window of interest should be considered as false alarms regardless of the error. We compensate this effect  by \textit{averaging} the above cost function over a sequence of FoV lenghts, $\setdeltau=\{\deltauone\le\dots\le\deltaul=\deltaudesired\}$.

%
%
%

\subsection{Design procedure}\label{sec:design-procedure}
Given a number of Tx and Rx elements, $\mtx$, $\nrx$, that can be placed in the available space $[\leftextremeTx,\rightextremeTx]$ and $[\leftextremeRx,\rightextremeRx]$, respectively, the optimal locations in our design methodology are given by the solution to the following problem:
\begin{subequations}\label{eq:opt-problem-design}
\begin{align}
\min_{\dtx\in\real^\mtx, \drx\in\real^\nrx} &\; \sum_{\Delta u\in\setdeltau}\tfrac{1}{\Delta u^2} \costdeltau
(\dtx, \drx) \label{eq:opt-problem-design-cost}
\\
\text{s.t.} 
&\;\; \dtxm{i+1}-\dtxm{i}>\mindistTx, \quad  i\in\until{\mtx-1} \label{eq:opt-problem-constraint_tx}
\\
 &\;\; \drxn{i+1}-\drxn{i}>\mindistRx,\quad i\in\until{\nrx-1} \label{eq:opt-problem-constraint_rx}
 \\
 &\;\; \dtxm{1}>\leftextremeTx,\quad \dtxm{\mtx}<\rightextremeTx \label{eq:opt-problem-constraint_tx_domain}
  \\
 &\;\; \drxn{1}>\leftextremeRx,\quad \drxn{\nrx}<\rightextremeRx \label{eq:opt-problem-constraint_rx_domain} ,
\end{align}
\end{subequations}
where $\mindistTx$, $\mindistTx$ codify the minimum separation between transmitters and receivers to avoid electromagnetic coupling
 and due to component size that depends on
 desired antenna power.


\begin{figure}[H]
	\includegraphics[width=0.87\linewidth]{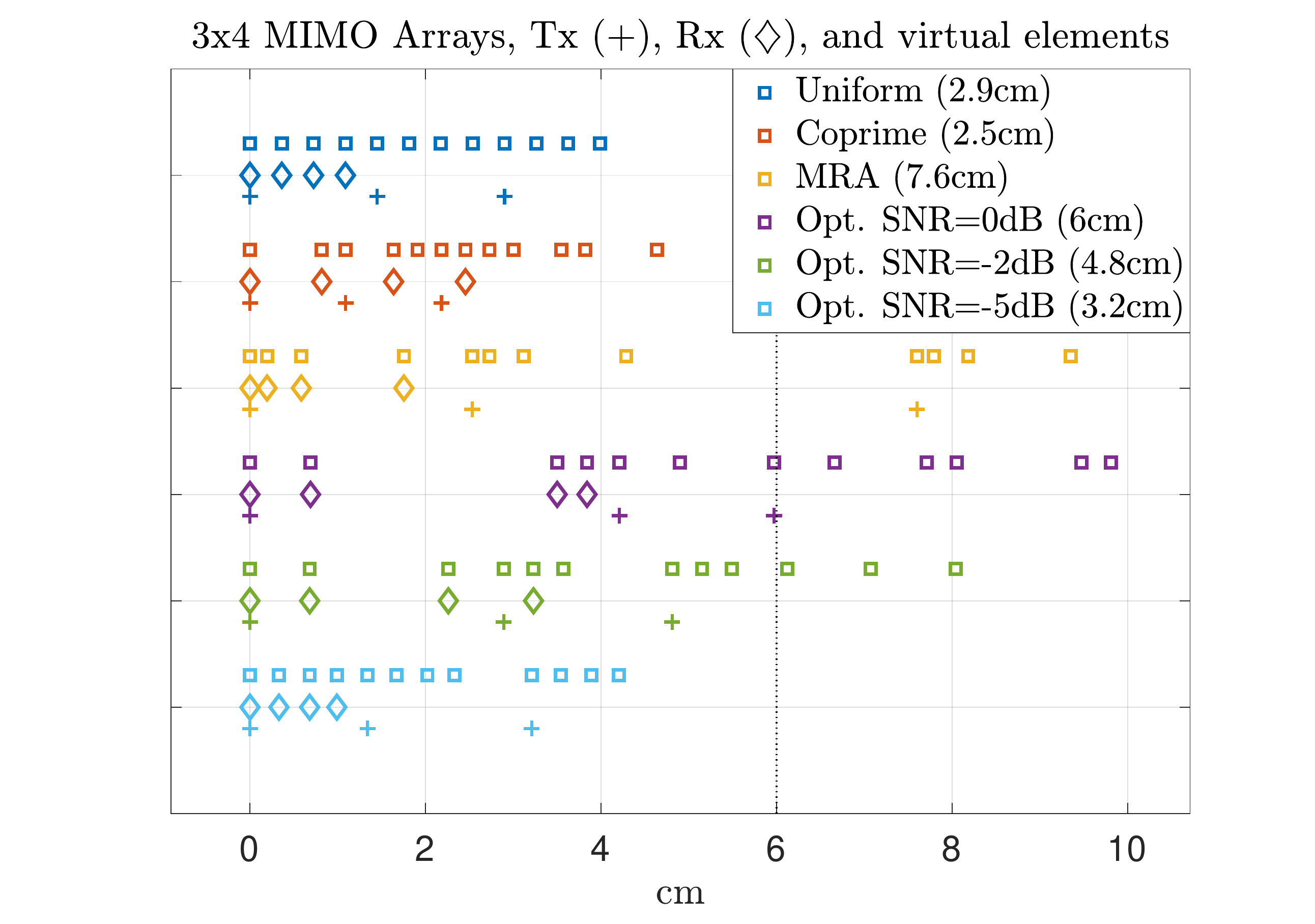}
	\caption{Optimized 3x4 MIMO arrays solving~\eqref{eq:opt-problem-design}. For reference, we include the uniform and coprime arrays, optimally dilated for the FoV $\pm 30^\circ$ with factors   $1.86$ and $1.4$, respectively, and also the MIMO MRA based on Cyclic Difference Sets~\cite{JD-QL-WG:09}. 
		%
	}
	\label{fig:array-geometries-3by4}
\end{figure}

\begin{figure}[bth]
	\includegraphics[width=0.99\linewidth]{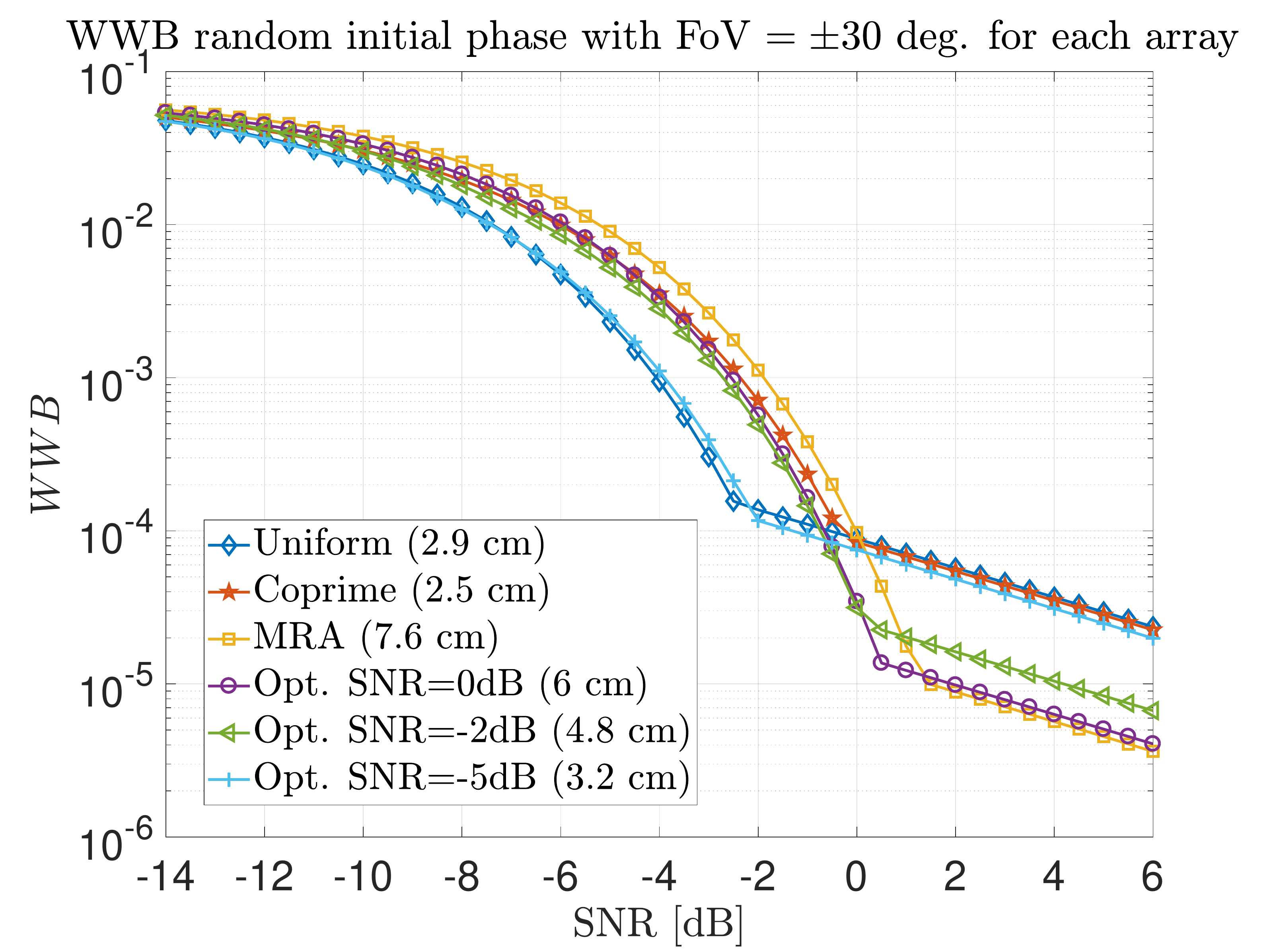}
	\caption{The WWB~\eqref{eq:opt-objective-design} versus SNR for the  arrays optimized using an average over FoVs $\pm (5^\circ, 15^\circ,30^\circ)$ in problem~\eqref{eq:opt-problem-design}.
	}
	\label{fig:wwb}
\end{figure}

%
%

\begin{figure*}[bth]
	\hspace*{-0.78in}
	\includegraphics[width=1.2\linewidth]{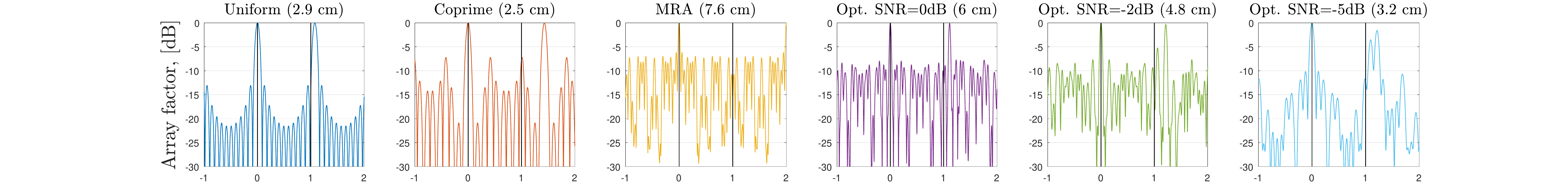}
	\caption{Array factor for the 3x4 MIMO arrays of Fig.~\ref{fig:array-geometries-3by4}, evaluated at $u-u_0=\sin(\phi)-\sin(\phi_0)$ for $\phi_0=0^\circ$ (centered), on $[-1, 1]$ and $\phi_0=-90^\circ$ (sided), on $[0, 2]$, both superimposed. We also plot the FoV length $[0, \deltau]$, with $\deltau=u_2-u_1 =\sin(30^\circ)-\sin(-30^\circ)=1$ (black lines). Note that the optimization \textit{pushes} high sidelobes and grating-lobes outside of the FoV.}
	\label{fig:beam}
\end{figure*}



We propose the following optimization strategy to solve problem~\eqref{eq:opt-problem-design}: i) The inner optimization over test points in~\eqref{eq:opt-objective-design} is solved using simulated annealing\footnote{Based on Matlab code by H\'ector Corte, available in MathWorks File Exchange, with some modifications for evaluation of vectorized functions in the inner optimization, and addition of constraints in the outer optimization.}~\cite{SK-CDG-MPV:83};
%
   ii) The outer optimization over antenna positions is solved  similarly but each candidate array needs to be feasible for the box constraints~\eqref{eq:opt-problem-constraint_tx_domain},~\eqref{eq:opt-problem-constraint_tx_domain}, and the convex constraints~\eqref{eq:opt-problem-constraint_tx},~\eqref{eq:opt-problem-constraint_rx}.
Although the number of antennas is higher than the dimension of test points, we find  suitable to invest enough computational effort in the inner optimization because failure to evaluate the optimal bound~\eqref{eq:opt-problem-design} results in optimistic predictions for candidate arrays and deterioration of the solution. In the Appendix, we exemplify for a given array that the number of local minima and steepness of the function depend strongly on the SNR.


Fig.~\ref{fig:array-geometries-3by4} shows examples of 3x4 arrays optimized with constraint parameters~$\mindistTx=3\lambda$, $\mindistTx=0.5\lambda$, with $\lambda=c/77$GHz$\approx3.9$mm, and $\rightextremeTx=\rightextremeRx=6$ cm. The cost function in~\eqref{eq:opt-problem-design-cost} is an average over FoVs of $\pm (5^\circ, 15^\circ,30^\circ)$. This has the effect of lowering the overall SLL, regardless of the distance to the mainlobe.
As expected, the aperture of the optimized arrays is larger when they are optimized for a higher SNR. At SNR$=0$, the optimization with the above setting uses the maximum available space.  
 The WWB~\eqref{eq:opt-objective-design} associated to these arrays, evaluated for the FoV$=\pm 30^\circ$, is depicted in Fig.~\ref{fig:wwb}. Note that the function employed in the optimization is the weighted average of the WWB over FoVs $\pm (5^\circ, 15^\circ,30^\circ)$.
 The WWB predicts the BMSE for a one-target model, and we observe that the new arrays present a lower value in relevant SNR intervals than the conventional arrays.



\section{DoA estimation Performance}\label{sec:doa-performance-validation}

In this section we compare the DoA estimation performance of the optimized arrays with 3x4 arrays that are commonly considered for the given spatial constraints, cf. Fig. \ref{fig:array-geometries-3by4}.  
%
%
For the comparison, we use a sparse reconstruction algorithm called FOCUSS~\cite{IFG-BDR:97,SFC-BDR-KE-KKD:05}.
\footnote{Code available online by Zhilin Zhang and David Wipf.} 
Briefly, sparse reconstruction refers to the problem of solving the under-determined system $y=Dx$, where $D\in\complex^{N\times k}$ has more columns than rows, $k>N$, under the assumption that $x$ is sparse. 
 In the application to DoA estimation, $y\in\complex^N$ is the $1$-snapshot measurement as in~\eqref{eq:measurement_model}, and $D$ is called a \textit{dictionary}, whose columns are the evaluations of the model~\eqref{eq:measurement_model} under the ideal noiseless case and with initial phase~$\varphi=0$, over a grid of hypotheses~$\{u_1,...,u_k\}$ within a FoV of interest.
  The solution, or sparse reconstruction $x\in\complex^{k}$, is expected to have nonzeros in the entries corresponding to the true DoA hypotheses. 
Each entry $i\in\{1,\dots,k\}$ of $x$ with a magnitude exceeding a threshold $\vert x_i\vert > \gamma$  constitutes a \textit{declared} target, or declaration, with DoA estimate $u_i$ associated to the corresponding column of the dictionary $D$.  We use $k=300$ in the FoV~$\pm 30^\circ$ 
and a threshold policy optimized for each array using a \textit{training} set of simulated scenarios in terms of the metrics of interest.
%
  %
  %
%



In the array evaluation, we use the following metrics: the Probability of Detection ($P_D$), the False Alarm Ratio (FAR), the Probability of Resolution ($P_R$), and the Root MSE (RMSE) of detections. The $P_D$, FAR, and $P_R$ are defined for each Monte Carlo realization, or trial, and are then averaged. In each trial we consider a detection window, DTW$=[-3^\circ, 3^\circ] $, around the realization of the ground-truth targets. If a declaration of the estimation algorithm falls inside a detection window, \textit{the corresponding target} is said to be \textit{detected}, otherwise, \textit{the declaration} is called a \textit{false alarm}.
The $P_D$ is defined, in each trial, as the quotient of the number of detections, divided by the total number of targets. The FAR\footnote{Different than the notion of \textit{False Alarm Rate} that considers the quotient over the number of grid points.} is consequently defined as the proportion of false alarms, i.e. the quotient of the number of declarations outside of any DTW around true targets, divided by the total number of declarations. In the case of two or several targets, $P_R$  is the proportion of times that all the targets are detected (see, e.g., in~\cite{AH-MGA-YDZ-FA:15}).
The accuracy of the detections is the standard RMSE \textit{restricted} to the declarations that are not false alarms.



\begin{figure*}[bth]

	\centering 
		\hspace*{-0.2in}
		 \begin{minipage}{1.05\textwidth}
	{\includegraphics[width=0.3267\linewidth]{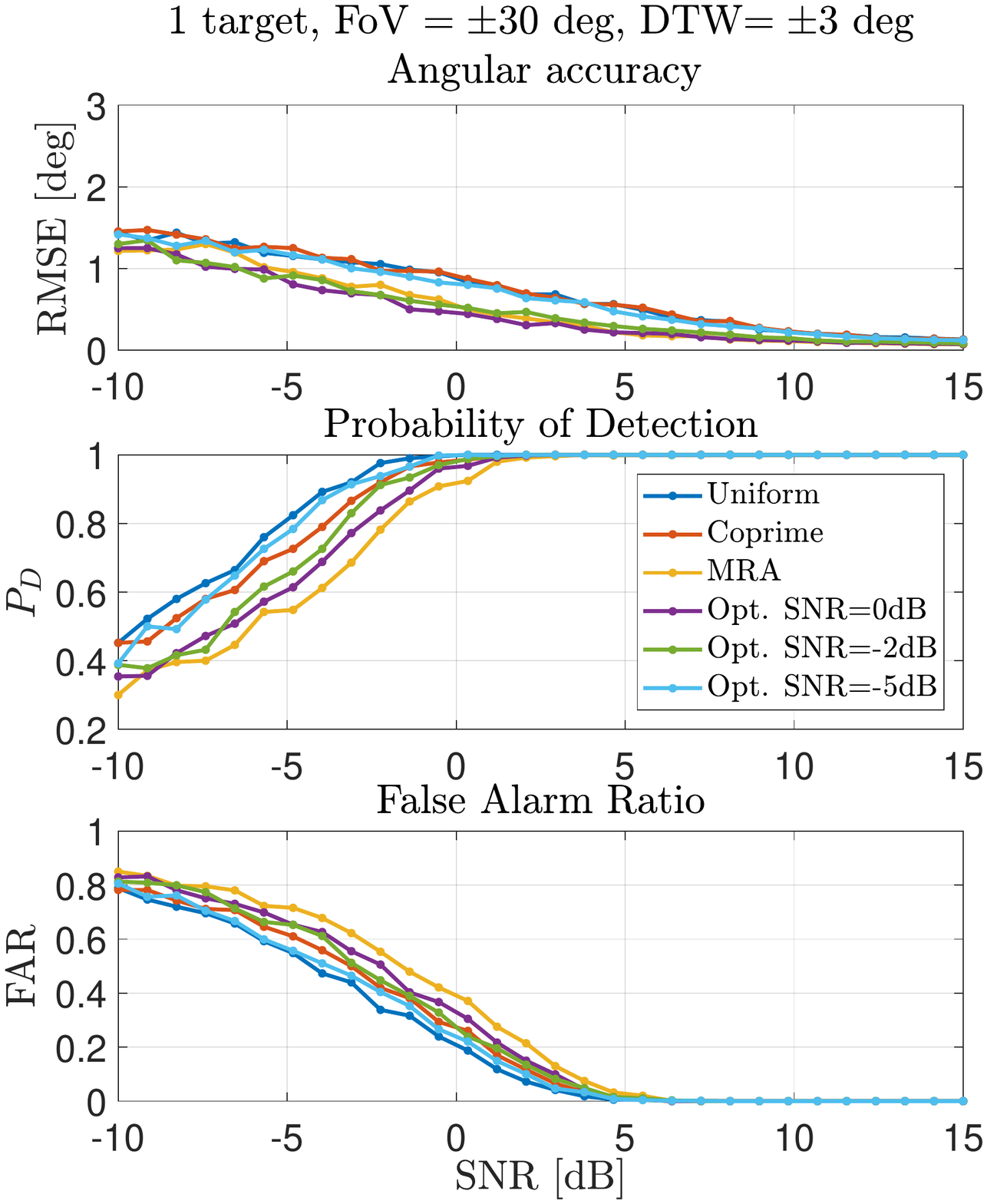}}
	{\includegraphics[width=0.3267\linewidth]{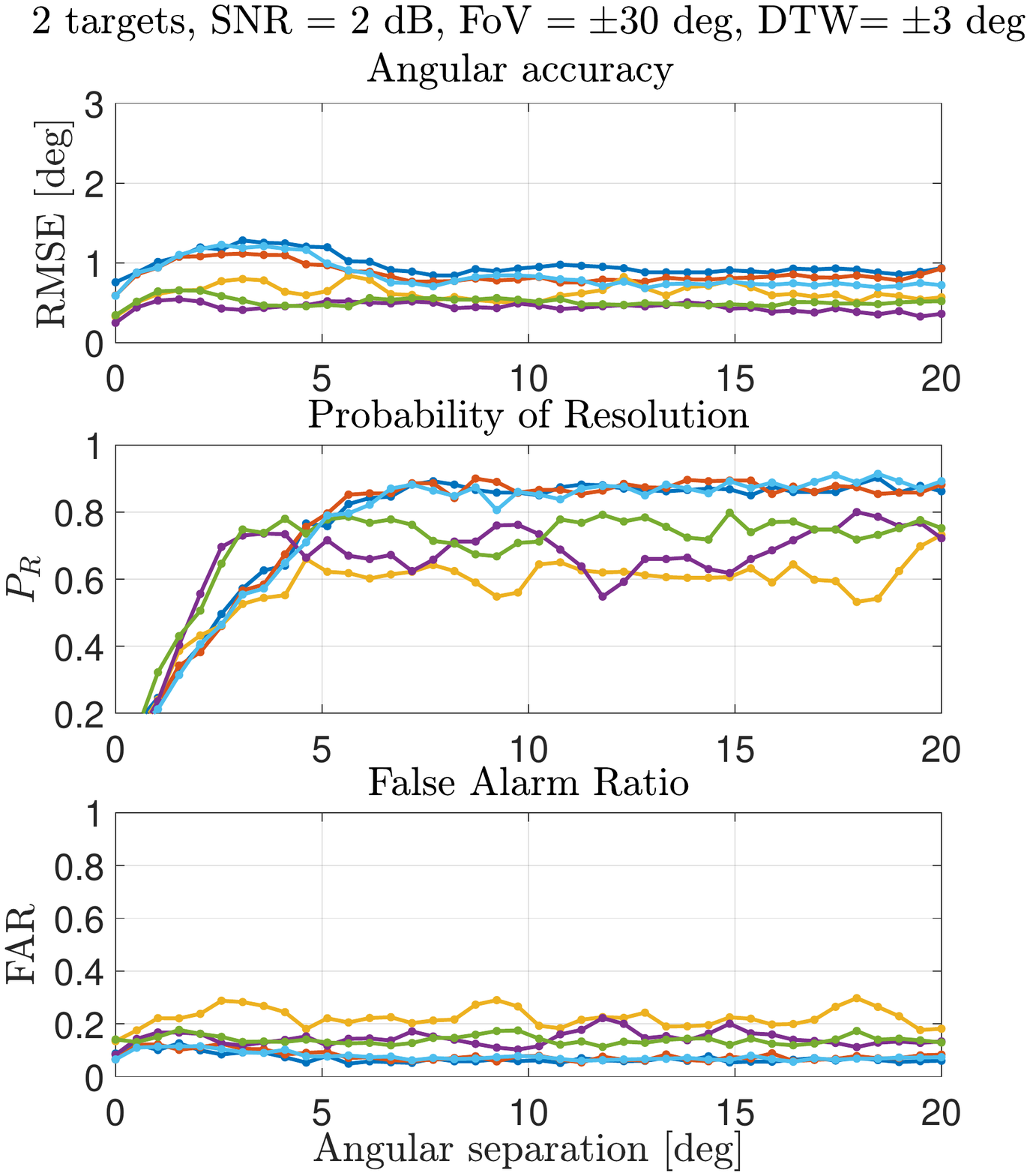}}
	{\includegraphics[width=0.3267\linewidth]{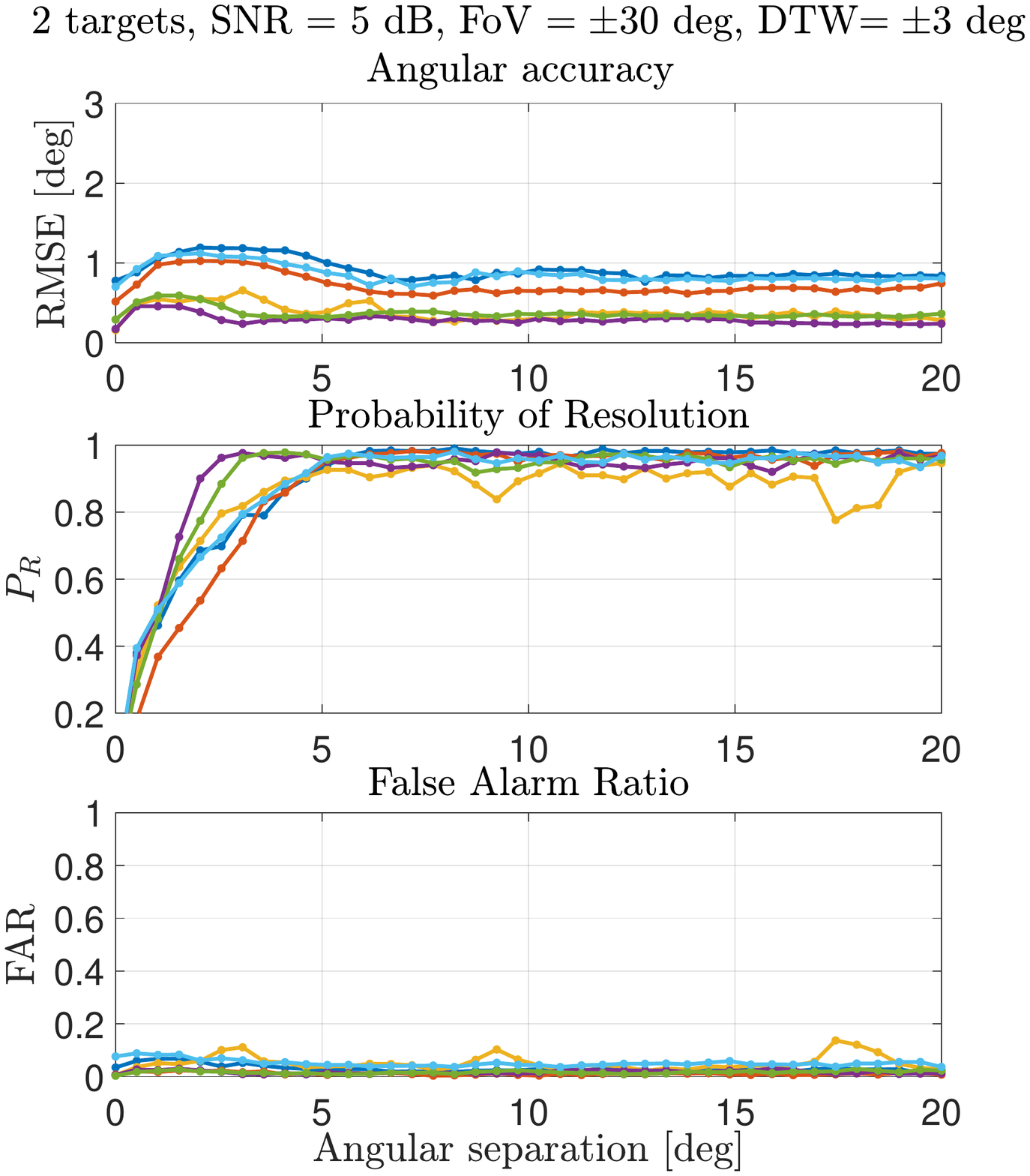}}
		 \end{minipage}
	\caption{Performance comparison of the 3x4 MIMO arrays in Fig.~\ref{fig:array-geometries-3by4} using FOCUSS algorithm with synthetic measurements. 
		 For one target, we represent the metrics versus SNR (left), and for two targets, versus angular separation for SNR$=2$ (middle) and for SNR$=5$ (right), computed with $500$ Monte Carlo realizations of the target positions for each SNR and angular separation.}
	\label{fig:cs-performance}
\end{figure*}


In Fig.~\ref{fig:cs-performance} we observe that, for one target, the optimized arrays improve on accuracy while keeping a higher $P_D$ and lower FAR than the MRA. 
The bigger benefit comes in terms of resolution for two targets. 
In this case, the optimized arrays achieve higher and more constant $P_R$ values than the MRA due to their overall lower sidelobe level (see Fig.~\ref{fig:beam}). 
With these results we show that for a 3x4 array and a FoV$=\pm 30^\circ$, using FOCUSS, the achievable resolution using the optimized arrays is about $2.5^\circ$,  while for the uniform dilated is about $5^\circ$, i.e., we can divide the angular resolution by two in the case of two targets with the same Radar Cross Sections (RCS).  



\section{Conclusions and future work}\label{sec:conclusions}


%

This work introduces a strategy to design optimal linear arrays for an arbitrary number of Tx and Rx antennas and desired spatial constraints. The WWB-based objective function includes as design choices the FoV and SNR values of interest, defining a trade-off between MLW and SLL based on a prediction of Bayesian MSE. This is in contrast with metrics that control directly the MLW and SLL. Resulting arrays, optimized for an average of FoVs, make more efficient use of the available space, and show an enhanced performance compared to conventional arrays, particularly the Probability of Resolution of two targets for a reduced FAR.
Future work includes generalizations to two-dimensional arrays, configurations with more antennas, and empirical analysis of resolution limits for two targets with different RCS.
%


\bibliographystyle{ieeetr} %

\appendix

\section{Derivation of the WWB used in our approach for array design}
First we introduce some preliminaries on the WWB, then discuss various DoA observation models, and finally construct the WWB used in this work for array design. 
\subsection{Preliminaries on the WWB}
Following \cite{HLVT-KLB:13}, chapter (4.4.1.4), the familiy of Weiss-Weinstein-Bounds (WWB) for an observation $y\in \complex^{N}$ and
 a vector parameter $\theta\in \real^{q}$, with joint probability density $p(y, \theta)$, is parametrized by the choice of \textit{test points}\footnote{In this work, the other class of parameters $S \defin [s_1, ..., s_M]\subset (0, 1)^M$ are chosen all equal to $1/2$. Note that~\cite{DTV-AR-RB-SM:14} found that this choice is optimal for linear arrays under their models.} $H \defin [h^1,..., h^M]\in \real^{q\times M}$
and obtained as
\begin{align}\label{eq:WWB}
\wwb(H) &= HQ^{-1}H\tp 
\end{align}
with the elements of the matrix $Q\in \real^{M\times M}$ given by
\begin{align}\label{eq:Q}
Q_{ij} &\defin \frac{E[(f_{y, \theta}(h^i)-f_{y, \theta}(-h^i))(f_{y, \theta}(h^j)-f_{y, \theta}(-h^j))]}{E[f_{y, \theta}(h^i)] E[f_{y, \theta}(h^j)]} ,
\end{align}
where the expectations are computed with the joint density $p(y,\theta)$, and
 we define $\map{f_{y, \theta}}{\real^q}{\realnonnegative}$ as
\begin{align}\label{eq:def-f-shorthand}
f_{y, \theta}(h) \equiv f(y; \theta+h, \theta) \defin \left(\frac{p(y, \theta+h)}{p(y, \theta)} \right) ^{\frac{1}{2}} .
\end{align}
All test points are valid as long as they are selected so that $Q^{-1}$ exists, which restricts the domain as follows,
\begin{align} \label{eq:validtestpoints}
\{H \defin [h^1,..., h^M]\in \real^{q\times M}\,:\, \Theta\cap (\Theta + h_m) \neq \emptyset,\, \forall m\},
\end{align} 
where $\Theta\defin \rm{supp}(p_\theta) \defin\{\theta \in \real^q: p_\theta(\theta) >0 \}$ denotes the support of the belief distribution on $\theta$. 
%
%
The WWB so defined
 lower bounds the Bayesian Mean Squared Error (BMSE) matrix of any estimator $\hat{\theta}(y)$, 
\begin{align*}
\Sigma \defin E[(\hat{\theta}(y) - \theta)(\hat{\theta}(y) - \theta)\tp]\succeq \wwb(H) ,
\end{align*}
in the sense of the Loewner-order (for Hermitian matrices $A\succeq B$ means that $A-B\succeq 0$, which is the notation for $A-B$ being positive-semidefinite), i.e.,
\begin{align*}
v^H\Sigma v \geq v^H \wwb(H)v
\end{align*}
for all vectors $v\in \complex^q$. In particular, if we are interested only in the estimation performance of a certain element of the parameter vector, e.g. $u = \theta_1\in \real$, we select $v = e_1$ (the first element of the canonical basis in~$\real^q$), and find a bound for the performance of $\hat{u}(y)\defin\hat{\theta}(y)_1$ in terms of the corresponding entry of the matrix $\wwb(H)$,
\begin{align}\label{eq:wwbDoA}
E[(\hat{u}(y)-u)^2]\geq \wwb(H)_{11} .
\end{align}
Next we discuss some observation models for DoA estimation and their consequences, which are captured by the WWB associated to the corresponding joint probability density $p(y,\theta)$.

\subsection{Conditional and unconditional models for DoA estimation}

Here we describe some modeling options of DoA estimation that can leverage the construction of the WWB presented in~\cite{DTV-AR-RB-SM:14}. We start with the general observation model
\begin{align}\label{mmodel}
y = A(\theta)s + w ,
\end{align}
where $A(\theta)\in \complex^N$ is a function of an unknown stochastic parameter vector $\theta\in\real^q$, and $s\in\complex$ can be considered deterministic or stochastic. Here $\noise\sim \mathcal{N}_\complex(0, \sigmanoise^2I_N)$ denotes white measurement noise with variance $\sigmanoise^2$ as in~\eqref{eq:measurement_model}. 
%
%
The \textit{conditional} model in~\cite{DTV-AR-RB-SM:14} assumes that $s\in\complex$ is deterministic and known, and allows, e.g. to identify $s$ with the transmitted waveform of the radar, which is of particular relevance for MIMO radar waveform design~\cite{WH-JT-RS:13,LXSFB08}.
%
%
Another usage (which we refer to as \textit{naive}) of the conditional model for DoA estimation assumes that $s = \absolute{ s } e^{i\varphi}$ is the target signal, in this case deterministic and known, while the steering matrix $A(\theta)\equiv A(u)= e^{idu}$ (for antenna positions  $d\in \real^N$ in units of $1/k=\lambda/(2\pi)$) depends only on the random DoA $u$. One inconvenience we encountered using this model is the dependence of optimized arrays on the choice of the array reference point used to define $d$, which we suspect to be caused especially by the modeling choice of known target phase $\varphi$. (In fact, the associated WWB depends in this case on the array factor through the real part $\Re\{ B(h_u) \}$, which has the consequence of modifying the condition for aliasing.)
%
%
%

In contrast, the \textit{unconditional} model in \cite{DTV-AR-RB-SM:14}, employed for DoA estimation, also identifies $s\in\complex$ in \eqref{mmodel} with the target signal, but it assumes it  obeys a zero-mean complex Gaussian distribution $s\sim \mathcal{N}_\complex(0, \sigmasignal^2)$. Stated equivalently, the target phase follows a uniform distribution $\varphi \sim U(-\pi, \pi)$, while the target magnitude $\absolute{s}$ is Rayleigh distributed (with scale parameter $\sigmasignal/\sqrt{2}$). Although this model seems appealing with regard to the phase, the Rayleigh distribution can make an undesired emphasis on target signals with low SNR values. As a consequence, arrays like the uniform have a competitive advantage for values of $\sigmasignal$ that are required to model a moderately high SNR regime in some applications where, in fact, sparse arrays are desirable.
This 
argues against the use of the unconditional model to design arrays for regimes of SNR that are relevant after range-Doppler processing for DoA estimation.

%
%

\begin{figure}[bth]
	\hspace*{-0.25cm}
	\includegraphics[width=1.05\linewidth]{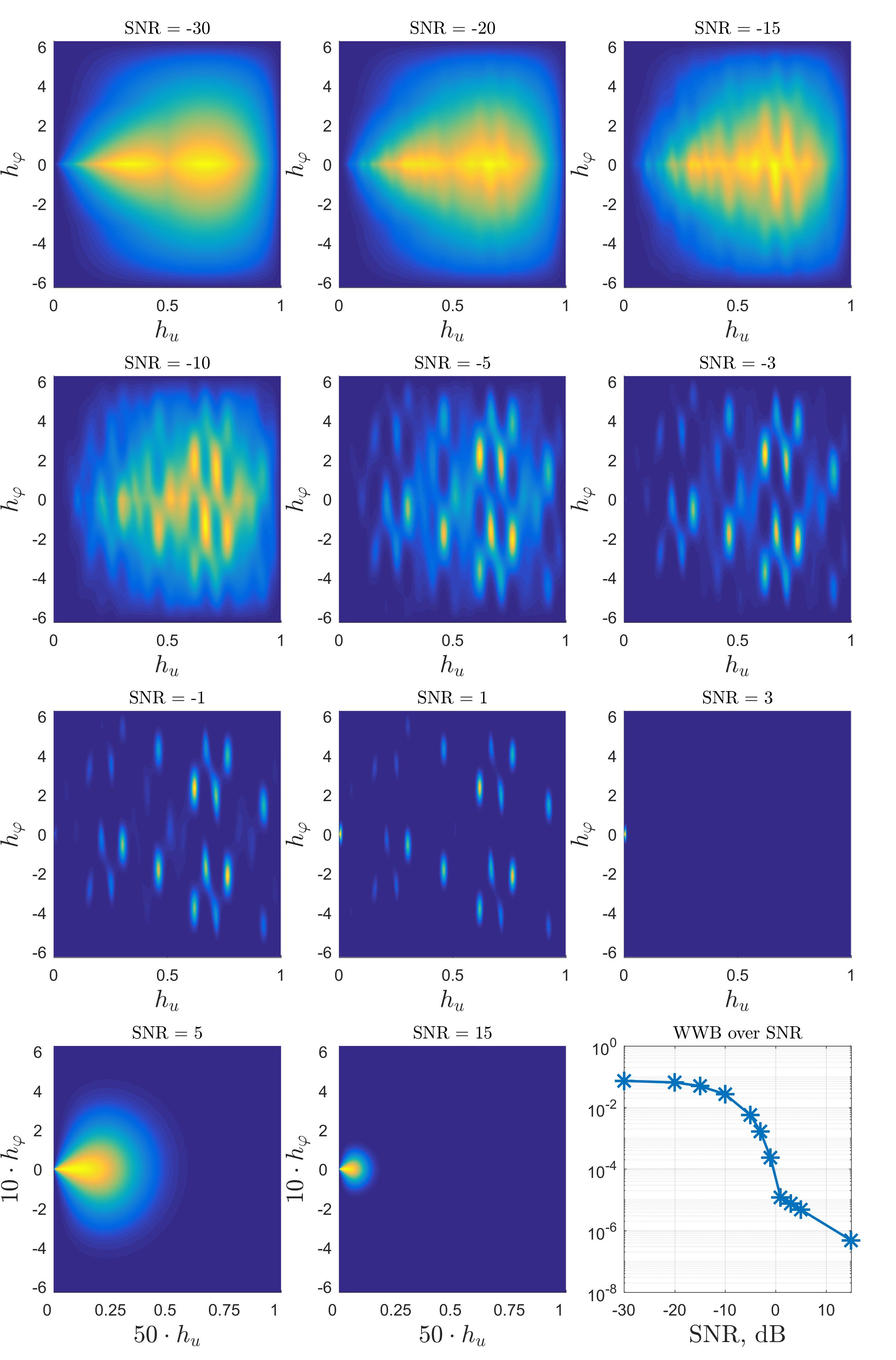}
	\caption{Surface plots over test points of $-WWB(-h_u, h_\varphi)$, defined in \eqref{eq:randomphaseWWB}, at different levels of SNR for a FoV of $\pm 30^\circ$ (i.e., $\Delta u = 1$), computed for the 3x4 MRA in Fig.~\ref{fig:array-geometries-3by4}. Note that for each candidate array, the negative function is minimized to compute the tightest bound in \eqref{eq:opt-objective-design}. (The color code is adapted to each plot. In addition, the axes for the plots at SNR $= 5$, and $15$ are scaled for better visualization because the WWB near the global maximum becomes steeper at higher SNR.) The plot to the southeast shows the maximum value of each previous plot over SNR.}
	\label{fig:WWBoverTP}
\end{figure}

\subsection{Construction of the WWB with random initial phase}\label{sec:wwb-derivation}
As a compromise between the unconditional and (naive) conditional model, we use a measurement model which assumes the target signal magnitude $\absolute{s}$ as deterministic and known, but regards the signal phase $\varphi$ as an uniformly distributed random variable in $(0,  2\pi)$. 
Denoting $\theta = [u, \varphi]\tp$, our measurement model in~\eqref{eq:measurement_model}, can be written as
\begin{align*}
y = A(\theta) \absolute{s} + w ,
\end{align*}
with $A(\theta) \defin e^{i\varphi}e^{idu}\in \complex^N$,
and prior distribution 
\begin{align}\label{eq:prior}
p_\theta(\theta) = \frac{1}{2\pi\Delta u}1_{(u_1, u_2)}(u)1_{(0, 2\pi)}(\varphi) ,
\end{align} 
where $1_{(a, b)}$ is the indicator function, which is $1$ in $(a,b)$ and $0$ otherwise,
and we recall that $\Delta u = u_2-u_1$ quantifies the radar's FoV. 
The measurement likelihood function is therefore given by the Gaussian density\footnote{We have included the phase explicitly in the parameter vector $\theta$ and thus obtained a multidimensional estimation problem instead of the alternative of employing likelihood functions $p(y\vert u) = \int_{-\pi}^{\pi} p(y \vert u, \varphi)p(\varphi)d\varphi$ involving the modified Bessel function of order zero, for which the evaluation of the WWB appears less tractable. 
}
\begin{align*}
p(y\vert \theta) = \frac{1}{\pi^N\sigma^{2N}}\exp\big(-\frac{\norm{ y - A(\theta)\absolute{s}}^2}{\sigma^2}\big) ,
\end{align*}
and the joint density needed for evaluation of the WWB can be obtained as
$p(y, \theta) = p(y\vert \theta) p_\theta(\theta)$.
Performing the calculations demanded by~\eqref{eq:WWB} in an analogue fashion to the treatment of the conditional model in~\cite{DTV-AR-RB-SM:14}, we obtain,\footnote{ 
	Expand the numerator of $Q_{ij}$ in \eqref{eq:Q}, express the joint density in terms of the likelihood and prior to separate integrations, use some null addition trick to solve the Gaussian integral over observations $y$, factor out the resulting expression noting that is independent of $\theta$, and employ the expression for the prior \eqref{eq:prior}.}
\begin{align*}
Q_{ij} = \frac{\eta(h^i, h^j)+\eta(-h^i, -h^j)-\eta(h^i, -h^j)-\eta(-h^i, h^j)}{\eta(h^i, 0)\eta(0, h^j)},
\end{align*}
where $h^i=[h^i_u, h^i_\varphi]\tp\in\real^2$ is a column test point, and we employ the following shorthand notations,
\begin{align*}
\eta(\mu, \rho) &\defin E_{p(y, \theta)}[f_{y, \theta}(\mu)f_{y, \theta}(\rho)] = \acute{\eta}_\theta(\mu, \rho) \frac{\vert \tilde{\Theta}(\mu, \rho)\vert}{2\pi \Delta u} ,
\\
\acute{\eta}_\theta(\mu, \rho) &\defin \exp( -\frac{cN}{2} [1-\Re \{e^{i(\rho_\varphi-\mu_\varphi)} B(\rho_u - \mu_u)\}]) ,
\\
B(h_u) &= \frac{\langle 1_N, e^{idh_u}\rangle} {\Vert e^{idh_u} \Vert^2}= \frac{1}{N} \sum_{n = 1}^N e^{id_nh_u} ,
\end{align*}
where we recall that $c \defin \absolute{s}^2  / \sigma^2\in\real$ is the SNR, $d\in\real^N$ codifies the antenna positions in units of $1/k=\lambda/(2\pi)$, 
and 
\begin{align*}
\vert \tilde{\Theta}(\mu, \rho) \vert \defin \vert \Theta\cap (\Theta+\mu) \cap (\Theta+\rho) \vert
\end{align*}
denotes the Lebesgue volume of the parameter-shifted support intersection, with $\Theta$ defined in~\eqref{eq:validtestpoints}. (Note that requirement~\eqref{eq:validtestpoints} ensures that $\eta(0, h_m)$ and $\eta(h_m, h_m)$ are nonzero.)

In this work we use a single-column test point
$H\equiv h \defin [h_u, h_\varphi]\tp\in \real^{2\times 1}$ for the WWB formulation in~\eqref{eq:WWB},\footnote{The authors are aware that this choice of test-point matrix conflicts with the condition $M\geq q$ suggested in~\cite{HLVT-KLB:13}. This condition is unnecessary for the derivation of the covariance inequality, and is only a necessary condition for a maximum-rank bound in~\eqref{eq:WWB}. We make the choice of rank-$1$ bound for the sake of computational efficiency and because it seems satisfactory for the purpose of having a tight lower bound in~\eqref{eq:wwbDoA}.} 
$\wwb(h) = \tfrac{1}{Q_{11}} h h\tp$.
With this choice, the final expression for the DoA component of the WWB as described in~\eqref{eq:wwbDoA} can be computed
as $\wwb(h_u,h_\varphi)_{11} =  \frac{h_u^2}{Q_{11}}$, yielding the expression~\eqref{eq:randomphaseWWB}.
%
%
%
%
%
To obtain a tight bound, we optimize the WWB~\eqref{eq:randomphaseWWB} over test-point values.\footnote{Other approaches exist: a common alternative chooses a fixed ``dense set" of test points related to the peaks of the array factor, as in~\cite{WH-JT-RS:13} for the Bobrovsky-Zaka\"i bound.} 
According to the condition in~\eqref{eq:validtestpoints}, the optimization is performed over
\begin{align*}
[h_u, h_\varphi] \in (0, \Delta u)\times (-2\pi, 2\pi),
\end{align*} 
where we have restricted the set according to a symmetry relation that can be derived from~\eqref{eq:randomphaseWWB}, namely, $\wwb(-h_u, h_\varphi)_{11} = \wwb(h_u, -h_\varphi)_{11}$, so that optimization over $(-\Delta u, 0)\times (-2\pi, 2\pi)$ is unnecessary. 
The optimal test points can be found with a global optimization algorithm like simulated annealing. Global optimization is necessary because this problem is in general nonconvex, as exemplified in Fig.~\ref{fig:WWBoverTP}, where we observe that the number of local minima and the steepness of the function strongly depend on the SNR.

%
%



\end{document}